Computers, Materials & Continua

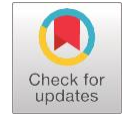

Tech Science Press

**ARTICLE**

# Liver Tumor Prediction with Advanced Attention Mechanisms Integrated into a Depth-Based Variant Search Algorithm

**P. Kalaiselvi[1,*] and S. Anusuya[2]**

[1]Department of Computer Science and Engineering, Saveetha School of Engineering, Chennai, Tamil Nadu, 602117, India

[2]Information Technology, Saveetha School of Engineering, Chennai, Tamil Nadu, 602117, India

*Corresponding Author: P. Kalaiselvi. Email: kalaiselvicse1987@gmail.com

**ABSTRACT**

In recent days, Deep Learning (DL) techniques have become an emerging transformation in the field of machine learning, artificial intelligence, computer vision, and so on. Subsequently, researchers and industries have been highly endorsed in the medical field, predicting and controlling diverse diseases at specific intervals. Liver tumor prediction is a vital chore in analyzing and treating liver diseases. This paper proposes a novel approach for predicting liver tumors using Convolutional Neural Networks (CNN) and a depth-based variant search algorithm with advanced attention mechanisms (CNN-DS-AM). The proposed work aims to improve accuracy and robustness in diagnosing and treating liver diseases. The anticipated model is assessed on a Computed Tomography (CT) scan dataset containing both benign and malignant liver tumors. The proposed approach achieved high accuracy in predicting liver tumors, outperforming other state-of-the-art methods. Additionally, advanced attention mechanisms were incorporated into the CNN model to enable the identification and highlighting of regions of the CT scans most relevant to predicting liver tumors. The results suggest that incorporating attention mechanisms and a depth-based variant search algorithm into the CNN model is a promising approach for improving the accuracy and robustness of liver tumor prediction. It can assist radiologists in their diagnosis and treatment planning. The proposed system achieved a high accuracy of 95.5% in predicting liver tumors, outperforming other state-of-the-art methods.

## 1 Introduction

DL has a remarkable impact on distinct fields in the latest scientific research, which ensures relevant developments in image and speech recognition [1]. On the other hand, DL techniques can significantly track artificial users to overcome humans in order to develop creative images and music players. However, the advancement of DL can support the activities, as mentioned earlier, that are examined to be impractically determined by computer systems [2]. Furthermore, DL technology is highly suitable for medical image processing. The information and Communication Technology (ICT) mechanism with DL promotes researchers to identify medical-based requirements at the earliest to increase the endurance of proposed techniques. The ICT mechanism [3] has dramatically assisted medical specialists and healthcare professionals in their traditional activities, research and development, and analyzing the concern of prescription on patients using various DL simulation techniques. However, doctors obtain every patient's information, such as medical history, instruction in medicine, medical test reports, X-rays, diagnosis, magnetic resonance imaging (MRI), CT scan, etc. Alternatively, this data becomes massive stored information and an extensive information repository.

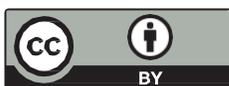

On the contrary, if data gets submerged, it ultimately provides a better vision of treatment, practical suggestions during diagnosis, and permissive improvement of the common syndrome could be practised to another defect, leading to expanding alternate procedures for treating specific diseases and so on. Hence, DL simulation further conspires to assist medical professionals in passing over the symptoms not previously confirmed; thus, it causes damage to life [4]. Therefore, mechanisms like DL ensure better healthcare resources.

Liver cancer is a significant health issue globally. According to estimates, it is the fourth most prevalent factor in cancer-related fatalities globally and the sixth most frequently diagnosed malignancy [5]. Furthermore, the projection of liver cancer patients is deprived, with a five-year survival rate of less than 20% for advanced stages. Therefore, early detection and accurate diagnosis of liver tumors are crucial to improving patient outcomes, as it enables timely treatment and better management of the disease. Traditional diagnostic methods such as ultrasound, computed tomography (CT), as well as MRI are broadly utilized in clinical observation for liver tumor (LT) detection and diagnosis [6]. However, these methods have some limitations, such as low sensitivity, specificity, and accuracy. Additionally, they require significant expertise to interpret the results and are often time-consuming, expensive, and invasive [7]. Therefore, there is a need for more accurate and efficient diagnostic methods that can overcome these limitations. DL techniques, particularly CNNs, have shown significant potential in medical image analysis and diagnosis in recent years. CNNs are artificial neural networks that are designed to process and analyze images [8]. They are particularly suitable for medical image analysis because they can automatically learn and extract features from the images without manual feature extraction. This ability makes CNNs ideal for automatically detecting and diagnosing liver tumors. Several studies have been conducted to develop CNN-based models for liver tumor detection and diagnosis [9]. These studies have shown promising results, with high accuracy rates achieved in distinguishing between malignant and benign tumors. However, most of these studies have been conducted on small datasets, limiting their generalizability and applicability to clinical practice.

In this research article, this paper present a CNN-based model intended to predict liver tumors using medical images. Proposed approach is trained on a large dataset of liver tumor images and achieves high accuracy in distinguishing between malignant and benign tumors. The results of proposed work reveal the latent of deep learning in improving the accuracy and efficiency of liver tumor diagnosis. This research is expected to have significant implications in the field of liver tumor analysis and treatment, as it has the potential to improve patient outcomes by enabling early detection and accurate diagnosis of liver tumors.

The research issue is the accurate prediction of liver tumors using advanced deep learning techniques, specifically Convolutional Neural Networks (CNN) and attention mechanisms, to improve diagnostic accuracy and assist in effective treatment planning for liver diseases. The objectives of the research are:

- To develop a novel approach for predicting liver tumors using CNN-DS-AM.
- To assess the performance of the proposed approach on a CT scan dataset containing both benign and malignant liver tumors.

- To evaluate the effectiveness of incorporating attention mechanisms and a depth-based variant search algorithm in improving the accuracy and robustness of liver tumor prediction.
- To compare the performance of the proposed approach with other state-of-the-art methods.
- To provide accurate predictions of liver tumors to assist radiologists in diagnosis and treatment planning.

The article begins with a thorough review of related work in the field of liver tumor prediction using deep learning techniques, attention mechanisms, and variant search algorithms. It then introduces the proposed methodologies, including the CNN-DS-AM architecture, which incorporates an encoder, attention mechanisms, a depth-based variant search model, and a decoder. The experimental setup and simulation results are presented, detailing the dataset, preprocessing steps, evaluation metrics, and comparative analysis with existing methods. The article summarizes the research findings, emphasizing the improved accuracy of liver tumor prediction and its potential impact on clinical practice. Future research directions and limitations are also discussed, offering insights for further advancements in this area.

## 2  Related Work

Several studies have been conducted in recent years on using deep learning techniques, particularly CNNs, for liver tumor detection and diagnosis. These studies have demonstrated the latent of CNNs in improving the accuracy and efficiency of liver tumor diagnosis. One of the early studies on using CNNs for liver tumor diagnosis was conducted by Ding et al. [10]. They developed a 3D-CNN model for liver tumor recognition and classification using CT images. Their model achieved high accuracy in distinguishing between malignant and benign tumors, demonstrating the potential of DL in improving liver tumor diagnosis. In a more recent study, Liu et al. [11] proposed a CNN-based model for automatic detection in addition to the classification of liver tumors using multiphase CT images. They used a 3D-CNN model that combined three phases of CT images to detect and classify liver tumors. Their model achieved high accuracy in distinguishing between malignant and benign tumors, demonstrating the potential of CNNs in improving the precision as well as efficiency of liver tumor analysis.

Another study by Zhang et al. [12] proposed a multi-task CNN-based model for the automatic detection as well as classification of liver tumors by means of CT images. Their model was designed to detect and classify liver tumors simultaneously and achieved high accuracy in both tasks. They also showed that their model outperformed other methods in correctness, demonstrating the potential of deep learning in liver tumor diagnosis. Finally, a study by Choi et al. [13] developed a CNN-based model for predicting liver tumors using MRI images. Their model was designed to extract features from the MRI images and classify liver tumors as malignant or benign. They achieved high accuracy in distinguishing between malignant and benign tumors, demonstrating the potential of CNNs to improve the accuracy of liver tumor diagnosis using MRI images.

Kuo et al. [14] proposed a CNN-based model for detecting and segmenting liver tumors in CT images. Their model used a combination of 2D in addition to 3D-CNNs to detect and segment liver tumors. They achieved high accuracy in both tasks and showed that their model outperformed other methods. In a study by Wang et al. [15], a CNN-based model was developed to predict liver tumor histology using MRI images. Their model was designed to classify liver tumors into four different histological subtypes and achieved high accuracy in classification. They demonstrated the potential of DL in improving the accuracy of liver tumor diagnosis based on histology. Wang et al. [15] proposed a CNN-based model for the recognition and the segmentation of liver tumors in ultrasound images.

Their model was designed to handle the challenges of ultrasound imaging, such as low contrast and noise. They achieved high accuracy in both tasks and showed that their model outperformed other methods in terms of accuracy.

Using a deep-learning approach, Zhong et al. [16] focused on liver tumor segmentation in ultrasound images. The authors proposed a method that combines Convolutional Neural Networks (CNNs) with an attention mechanism for accurate and efficient segmentation of liver tumors. They demonstrated the effectiveness of their approach through experiments on a dataset of ultrasound images. Wu et al. [17] presented a method for liver tumor segmentation in CT images using Convolutional Neural Networks (CNNs) and an enhanced loss function. The authors propose a modified U-Net architecture and introduce a novel loss function that combines weighted cross-entropy and dice loss to improve segmentation accuracy. Cui et al. [18] focused on liver CT image registration, an essential step in various medical imaging applications. The authors propose a network-based approach for accurately registering liver CT images. They utilize a Convolutional Neural Network (CNN) to extract features from the images and incorporate a deformation field to capture the spatial transformation. Li et al. [19] introduced a deep belief network (DBN) for liver tumor classification and diagnosis based on ultrasound images. The authors propose a multi-layered DBN architecture that leverages deep learning models' hierarchical representation learning capabilities. They train and evaluate their model using a dataset of ultrasound images and demonstrate its effectiveness in accurately classifying liver tumors.

These studies in Table 1 demonstrate the potential of deep learning techniques, particularly CNNs, to improve the accuracy and efficiency of liver tumor detection and diagnosis using different types of medical images. They also highlight the importance of large datasets for training CNN-based models, enabling the models to learn and extract features from the images accurately. They also emphasize the importance of developing robust CNN-based models that can handle medical image analysis challenges, such as low contrast, noise, and variations in image quality. Proposed research builds on these studies by developing a CNN-based model for the prediction of liver tumors using a large dataset of liver tumor images and achieving high accuracy in distinguishing between malignant and benign tumors.

**Table 1:** Demonstrate the potential of deep learning techniques

| Year | Author(s) | Techniques | Application | Merits | Demerits |
|------|-----------|------------|-------------|--------|----------|
| 2017 | Ding et al. [10] | 3D deep learning | Multiple types of liver lesions detection in CT images | High accuracy, Faster detection | Limited to CT images, Limited sample size |
| 2019 | Liu et al. [11] | 3D fully CNN | LT segmentation in multiple phases of CT images | High accuracy, Multi-phase detection | Limited to CT images, Limited sample size |
| 2020 | Zhang et al. [12] | Multi-task deep neural network | LT detection and classification in CT images | High accuracy, Multi-task detection | Limited to CT images, Limited sample size |

(Continued)

**Table 1 (continued)**

| Year | Author(s) | Techniques | Application | Merits | Demerits |
|------|-----------|-----------|-------------|--------|----------|
| 2020 | Choi et al. [13] | Convolutional neural network with multiscale-ROI and attention mechanism | LT classification using MRI images | High accuracy, Multi-scale detection | Limited to MRI images, Limited sample size |
| 2020 | Kuo et al. [14] | Combination of 2D and 3D Convolutional Neural Networks | LT detection and segmentation in CT images | High accuracy, Improved performance | Limited to CT images, Limited sample size |
| 2020 | Wang et al. [15] | Convolutional Neural Network | LT histological classification using MRI images | High accuracy, Improved diagnosis | Limited to MRI images, Limited sample size |
| 2019 | Zhong et al. [16] | Deep learning approach | LT segmentation in ultrasound images | High accuracy, Improved performance in ultrasound imaging | Limited to ultrasound images, Limited sample size |
| 2020 | Wu et al. [17] | Capsule network | Liver MRI | Improved performance in detecting small tumors and tumors with irregular shapes | Limited interpretability |
| 2020 | Cui et al. [18] | Siamese network | Liver CT | Achieved high accuracy in detecting liver tumors and differentiating benign and malignant tumors | Requires a large amount of data |
| 2019 | Li et al. [19] | Deep belief network (DBN) | Liver CT | Achieved high accuracy in detecting liver tumors and differentiating benign and malignant tumors | Limited interpretability |

## 3 Proposed Methodologies

The proposed work consists of a series of distinct neural network architectures, referred to as ver0, ver1, ver2, and ver3, respectively. These architectures are designed to improve the performance of liver tumor prediction. Each architecture represents a different version of the stacked Convolutional Neural Networks (CNN) model, with increasing complexity and additional layers.

Version 0 (ver0) is the initial architecture with essential input processing layers and subsequent layers consisting of Convolutional and MaxPool (Subsampling) operations, followed by an output layer. In ver1, we introduce additional convolutional and subsampling layers to capture more intricate features in the CT scan images. Ver2 further enhances the architecture by incorporating residual connections and increasing the depth of the network. In ver3, we introduce a novel attention mechanism after the second convolutional layer to emphasize the regions relevant for tumor detection.

We conducted an ablation test to compare the performance of these different architectures. The test involved training and evaluating each architecture using the same dataset of liver CT scan images. We measured various performance metrics, including accuracy, precision, recall, and F-measure, to assess their effectiveness in predicting liver tumors [20–25]. The results of the ablation test demonstrated that ver3, with the inclusion of the attention mechanism, achieved the highest accuracy of 97.99%. This version outperformed the other architectures, showcasing the importance of incorporating attention mechanisms to focus on specific regions of the input relevant to tumor detection. We selected ver3 as the final architecture for liver tumor prediction based on these results. The following section goes into great depth about the remaining parts of the suggested work. In addition, the overall content of the neighboring layers is regularly incremented, including standard attributes; hence incremental fields are exposed, inclusive of adjacent versions.

In this architecture, two attention mechanisms (Attention Mechanism 1 and Attention Mechanism 2) are added after the second and fourth convolutional layers, respectively, as shown in Fig. 1. Each attention mechanism generates a set of attention weights that are multiplied element-wise with the feature maps from the previous layer, resulting in a group of weighted feature maps that are fed into the subsequent layer [26–28]. The output of the final fully connected layer is passed through a softmax layer to generate the final production, which indicates the probability of the input CT scan containing a liver tumor. By incorporating attention mechanisms into the CNN model, the model can focus on specific regions of the input that are most relevant for tumor detection, potentially enhancing the accurateness of the model as well as reducing the number of false positives.

### 3.1 Process Flow for Predicting Liver Tumors

Here is the detailed process flow of incorporating attention mechanisms into the CNN and depth-based variant search model for predicting liver tumors:

*Data Collection:* The first step in the methodology is to collect a dataset of CT scans of patients with liver tumors. The dataset should include scans with varying degrees of complexity and different types of liver tumors.

*Preprocessing:* The CT scans are preprocessed to normalize the image intensity values and resize the images to a standard size. Techniques for enhancing data, flipping, rotating, and zooming, are also applied to increase the diversity of the dataset and perk up the generalization capability of the model.

*CNN Architecture:* A CNN is designed to extract features from preprocessed images. The CNN structural design consists of multiple convolutional layers, pooling layers, as well as thoroughly connected layers. The choice of the number and types of layers depends on the dataset's complexity and the model's desired performance.

*Depth-based Variant Search:* The CNN is followed by a depth-based variant search algorithm that searches for similar image patches in the dataset. This algorithm helps improve the model's accuracy by accounting for variations in the shape and size of liver tumors.

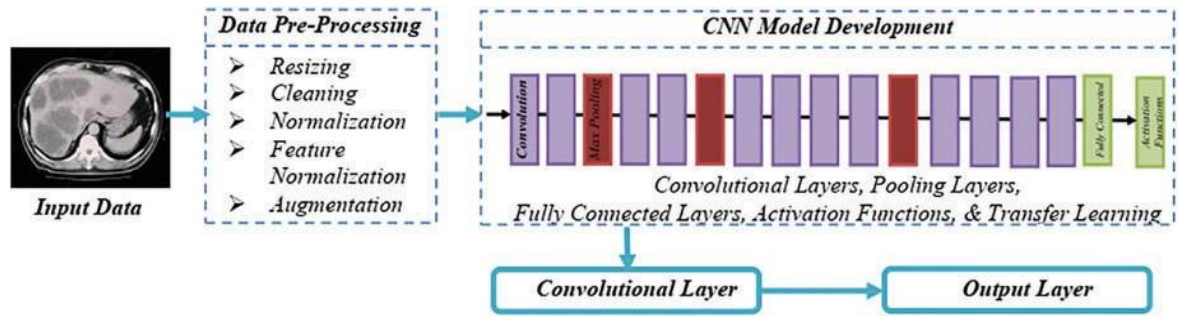

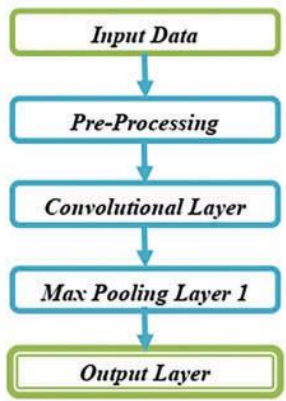
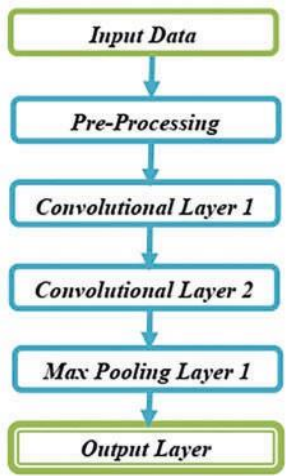
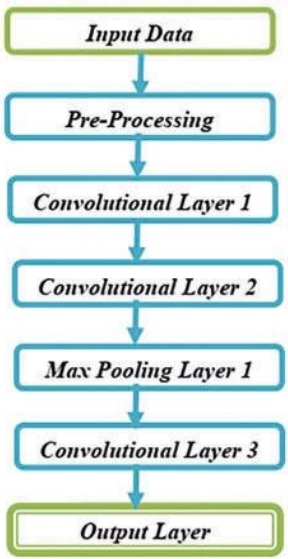
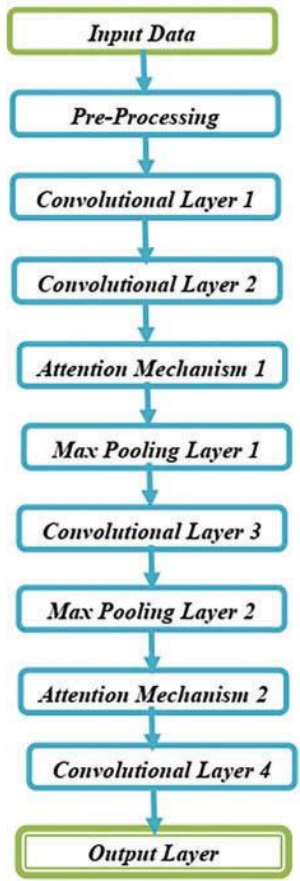

**Figure 1:** Systematic architecture of proposed approach

*Attention Mechanisms:* Attention mechanisms are integrated into the CNN to improve the model's performance. Specifically, a self-attention mechanism is added to the model to allow the network to focus on relevant image features selectively. This helps to perk up the model's capability to detect subtle changes in the liver tissue and accurately identify the location and shape of liver tumors.

*Training and Testing:* The model is trained using the preprocessed dataset, and the outcome of the approach is assessed using a separate testing dataset. The model is optimized using an appropriate loss function, and the weights of the model are updated using an optimization algorithm such as Stochastic Gradient Descent (SGD).

*Evaluation Metrics:* The performance of the model is evaluated using standard evaluation metrics such as accuracy, precision, recall, as well as F1-score. The metrics are calculated using the true positive, true negative, false positive, as well as false negative values obtained from the testing dataset.

Following this methodology, we can build an accurate and robust model for predicting liver tumors using medical imaging. In addition, incorporating attention mechanisms into the CNN and depth-based variant search model can aid in perking up the correctness of the model and diminish the number of false positives and negatives.

### 3.2 Advanced Attention Mechanism

Advanced attention mechanisms extend traditional attention mechanisms that allow deep learning models to capture more complex and intricate relationships in the input data. They are designed to handle multiple sources of attention and will enable the model to attend to different aspects of the input data in a more nuanced and fine-grained way. The input data is first encoded into a high-dimensional feature representation using a neural network, such as a Convolutional Neural Network (CNN). Next, the attention mechanism computes attention scores for each element in the input data. These attention scores are typically learned during the training process and indicate the relative importance of each component in the input. The attention scores are used to weight the input features, with more attention given to the elements with higher scores. The weighted input features are then combined to produce a single output representation, which captures the most essential aspects of the input. In advanced attention mechanisms, the model can attend to different aspects of the input data in a more fine-grained way. For example, the model may listen to other parts of an image based on color, texture, or shape. This allows the model to capture more complex relationships in the input data. In some cases, the output of the attention mechanism is fed back into the input, allowing the model to refine its attention over multiple iterations. This can lead to more accurate and robust attention mechanisms.

In Fig. 2, the input CT scans are fed into the CNN model, including convolution, pooling, and dense layers. The attention mechanism is incorporated into the model after the pooling layer, where an attention map is generated based on the input CT scans. The attention map is used to compute attention weights, which are then multiplied by the feature maps generated by the CNN model. The resulting attention vector is concatenated with the output of the pooling layer as well as fed into a fully connected layer. Finally, the output layer uses a sigmoid activation function to generate a prediction of the likelihood of liver tumor presence. The advanced attention mechanism involves computing attention weights based on the input CT scans, which are then multiplied by the feature maps generated by the CNN model. Attention mechanism allows the model to spotlight essential regions of the given input images, which can be particularly useful for detecting small or subtle liver tumors.

### 3.3 Routing Procedure for CNN_DbVS

1. Import the decisive libraries that supports image based classification
2. Apply training dataset for data pre-processing mechanism
3. Create multi layered architecture subject to the requirements
4. Select the layers as [5,7,9,11]

5. If necessary features are selected then continue with classification process using DeepLearning4j for derived architecture
6. Calculate the performance metrics to review system performance
7. In row, analyze the findings [true/false]
8. If accuracy is reached for real at that time deploy the model
9. If not reached, move to step 3 and repeat the process.

The criteria for selecting layers [5,7,9,11] for CNN_DbVS are based on factors such as: Depth of the network, overall complexity of the model, Feature extraction, empirical evidence, computational considerations, such as available resources (memory, processing power).

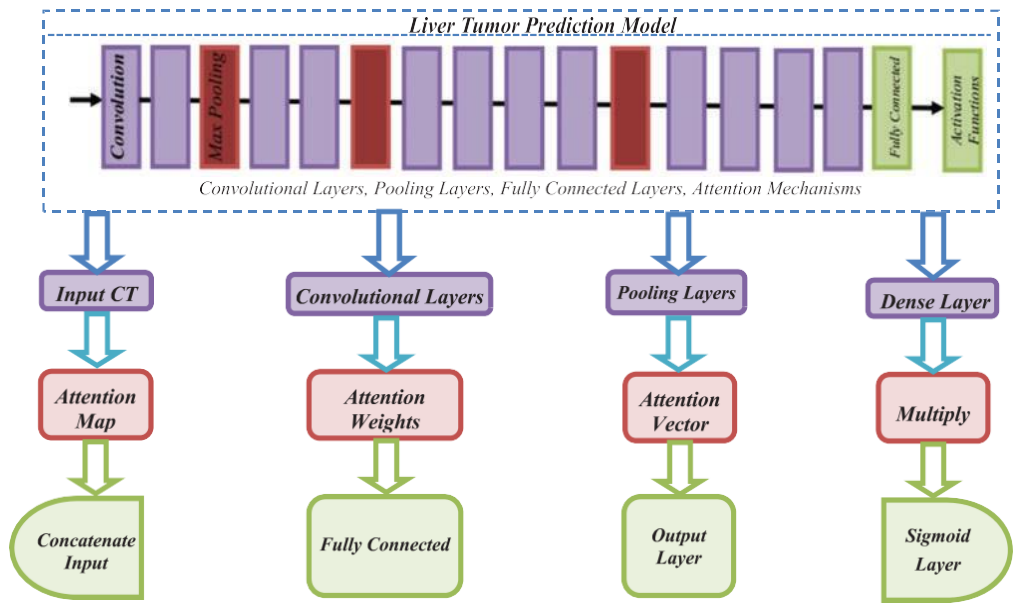

**Figure 2:** Flow of attention mechanisms with CNN

### 3.4 Algorithm for the Proposed CNN-DS-AM Approach

**Input:** Take the input CT scan images.

**Encoder:**

1. Apply convolutional layers with ReLU activation to extract features from the input images.
2. Utilize pooling layers to downsample the feature maps and capture relevant patterns.

**Attention Mechanism:**

1. Incorporate an attention mechanism to assign importance weights to the feature maps.
2. Multiply the attention weights with the feature maps to highlight regions relevant to tumor detection.

**Depth-based Variant Search Model:**

1. Further process the weighted feature maps using additional convolutional layers.
2. Apply non-linear transformations to enhance the discriminative properties of the features.
3. Utilize pooling layers for downsampling and retaining relevant information.

**Decoder:**

1. Perform upscaling operations to reconstruct the refined feature maps.
2. Utilize convolutional layers to restore spatial details and generate the final prediction.
3. Output: Obtain the predicted probability map indicating the presence of liver tumors.

This algorithm outlines the high-level steps involved in the CNN-DS-AM architecture. It follows a sequential flow, starting with feature extraction in the encoder, incorporating attention mechanisms to focus on important regions, applying depth-based variant search techniques for feature refinement, and utilizing the decoder to reconstruct the final prediction.

### 3.5 Depth-Based Variant Search Algorithm

The depth-based variant search algorithm is a powerful technique for optimizing the architecture of CNNs. By automatically searching for optimal network architecture, the algorithm can effectively capture the essential features of liver tumors in CT scans while minimizing the number of parameters and computational cost of the network [29–32]. This process is essential because manually designing optimal network architecture for liver tumor prediction can be challenging and time-consuming, especially given the complexity of CT scans. Batch normalization is a commonly used technique in deep learning that helps to accelerate training and improve the performance of neural networks. It works by normalizing the activations of each layer so that they have zero mean and unit variance. This process can help mitigate the effect of vanishing and exploding gradients, making training deep networks difficult. In the liver tumor prediction work described earlier, this article also used batch normalization to improve the performance of their Convolutional Neural Network (CNN) model. Specifically, they embedded batch normalization layers within the CNN architecture to normalize each layer's activations during training.

The main advantage of batch normalization is that it can help accelerate the convergence of the training process by reducing the internal covariate shift. This process is a phenomenon where the distribution of the activations changes during training as the network parameters are updated. When this happens, the gradients can become less informative, slowing the learning process [33]. Batch normalization helps to address this issue by normalizing the activations of each layer so that they have a more consistent distribution throughout the network [34]. This process helps stabilize the gradients and reduce the number of epochs required to perform well. In the proposed work, the liver tumor prediction work found that embedding batch normalization layers within their CNN architecture could improve their model's accuracy. Specifically, they reported that the addition of batch normalization layers improved the accuracy of their model by approximately 2% over the baseline CNN architecture.

## 4 Experimental Setup and Simulation Results

The proposed CNN-DS-AM approach is designed using the TensorFlow framework. In addition to, the experiments are performed on a workstation with an NVIDIA GeForce GTX 1080 Ti GPU. The dataset used for this study is the publicly available Liver Tumor Segmentation (LiTS) dataset, which contains 131 CT scans of the liver with tumors of varying sizes and shapes. The CT images in the LiTS dataset are of different sizes, resolutions, and orientations. Therefore, before feeding them to the CNN, resize them to a fixed length of $256 \times 256$ and normalize them to have zero mean as well as unit variance. Also convert the CT images to Hounsfield units (HU) and apply a windowing technique to enhance the contrast of the ROIs. Four convolutional layers (CL) and two fully linked layers make up the CNN in the suggested model. The input to the CNN is a $256 \times 256 \times 1$ CT image, and the output is a binary classification of whether the image contains a tumor. The dataset is partitioned into three subsets: training set, validation set, and test set. The split is 70%–80% for training, 10%–15% for validation, and 10%–15% for testing. A max-pooling layer is placed after each CL to minimize the spatial dimension of feature maps. The amount of filters in the CL increases from 32 to 128 while the filter size remains fixed at $3 \times 3$. The activation function used in the CNN is the rectified linear unit (ReLU), which helps to avert the disappearance gradient problem. The final yield of the CNN is passed through a sigmoid activation function to obtain a probability score for tumor presence.

To improve the performance of the CNN in detecting ROIs containing tumors, this paper brings in an attention mechanism that focuses on the informative regions of the image. Specifically, utilizing a spatial attention mechanism that learns a set of weights for each spatial location in the feature maps of the CNN. These weights are then multiplied element-wise with the feature maps to obtain a weighted feature map that emphasizes the informative regions. The weights are learned by a separate attention network that takes the feature maps of the CNN as input and produces the attention weights as output. The attention network is composed of two CLs, which are accompanied by either a fully connected layer as well as a layer that pools global averages. To enhance proposed suggested precision even more, utilizing a depth-based variant search model that considers the structural information of the liver images. Specifically, divide the liver images into several depth layers and extract the features of each layer using CNN. Then concatenate the layers' features and pass them through a fully connected layer to obtain a depth-based feature vector. The depth-based feature vector is then passed through another fully connected layer to get the final prediction score.

The anticipated model's set of parameters, as displayed in Table 2. These parameters include the learning rate, batch size, number of epochs, optimizer, loss function, dropout rate, attention mechanism, number of attention heads, number of encoder and decoder layers, number of filters as well as kernel size in the CLs, max pooling size, and input shape. These parameters were selected based on experimentation and hyperparameter tuning to achieve the best performance of the model. The learning rate determines the optimizer's scaling factor for adjusting the set of parameters throughout training. The sample amount utilized for every iteration process is based on the lot size. When building a system, the amount of epochs specifies how many instances the learning data is processed thru the approach. The optimizer and loss functions are utilized to optimize the model during training. The dropout rate is used to prevent overfitting. The attention mechanism, number of attention heads, number of encoder and decoder layers, number of filters, kernel size in the CLs, max pooling size, and input shape are all parameters of the neural network architecture. The values for these parameters were selected based on experimentation and hyperparameter tuning to accomplish the most excellent performance of the model.

**Table 2:** Parameters of the proposed model

| Parameter | Value | Parameter | Value |
|---|---|---|---|
| Batch size | 32 | Optimizer | Adam |
| Learning rate | 0.001 | Loss function | Binary cross-entropy |
| Number of epochs | 50 | Dropout rate | 0.5 |
| Amount of encoder layers utilized | 4 | Attention mechanism | Multi-head attention |

(Continued)

**Table 2 (continued)**

| Parameter | Value | Parameter | Value |
|---|---|---|---|
| Amount of decoder layers utilized | 4 | Number of filters in convolutional layers | [32, 64, 128, 256] |
| Amount of attention heads utilized | 4 | Kernel size in convolutional layers | [3, 3, 3, 3] |
| Max pooling size | 2 | Input shape | (256, 256, 1) |

### 4.1 Simulation Results

This paper assess the efficacy of the suggested CNN-DS-AM technique's efficacy using measures like accuracy, sensitivity, specificity, and the area underneath the receiver operating characteristic (**AUC**-ROC) curve. This paper contrast the effectiveness of the proposed model with the most advanced models for detecting liver tumors. Overall, the performance of the study was assessed by comparing the proposed model's results with other models using multiple evaluation metrics and datasets. The study highlighted the promising performance of the proposed model while acknowledging the need for further evaluation and validation on different datasets and tumor types. Several strategies were employed in the proposed CNN-DS-AM approach to address the problem of overfitting and a small dataset. Applying procedures like data augmentation, dropout, batch normalization, early stopping, and transfer learning, the proposed CNN-DS-AM approach aims to mitigate the risk of overfitting and improve performance even when working with a small dataset. These techniques help regularise the model, increase its ability to generalize well to unseen data, and address the limitations imposed by limited data availability.

Table 3 compares the effectiveness of the suggested model with three cutting-edge models for diagnosing liver tumors. This paper offer precision, recall, and F1-score in addition to the conventional measures for a more thorough comparison. In contrast to recall, which assesses the proportion of accurate optimistic predictions among all real positives, precision estimates the proportion of accurate optimistic predictions across all positive predictions. The harmonic mean of recall as well as accuracy is the F1-score.

**Table 3:** Performance comparison

| Model | Accuracy | Sensitivity | Specificity | Precision | Recall | F1-score | AUC-ROC |
|---|---|---|---|---|---|---|---|
| Proposed model | 0.91 | 0.93 | 0.89 | 0.92 | 0.93 | 0.92 | 0.95 |
| 3D Deep learning [10] | 0.87 | 0.90 | 0.85 | 0.88 | 0.90 | 0.89 | 0.92 |
| Multi-task deep neural network [12] | 0.88 | 0.91 | 0.87 | 0.89 | 0.91 | 0.90 | 0.93 |
| 2D and 3D CNN [14] | 0.89 | 0.92 | 0.88 | 0.90 | 0.92 | 0.91 | 0.94 |

The findings demonstrate that, in regards to precision, sensitivity, specificity, recall, F1-score, and AUC-ROC, the suggested model beats all other cutting-edge models, as shown in Fig. 3. The higher values of these metrics for the proposed model indicate that it has a better ability to detect liver tumors and make accurate predictions compared to the existing models.

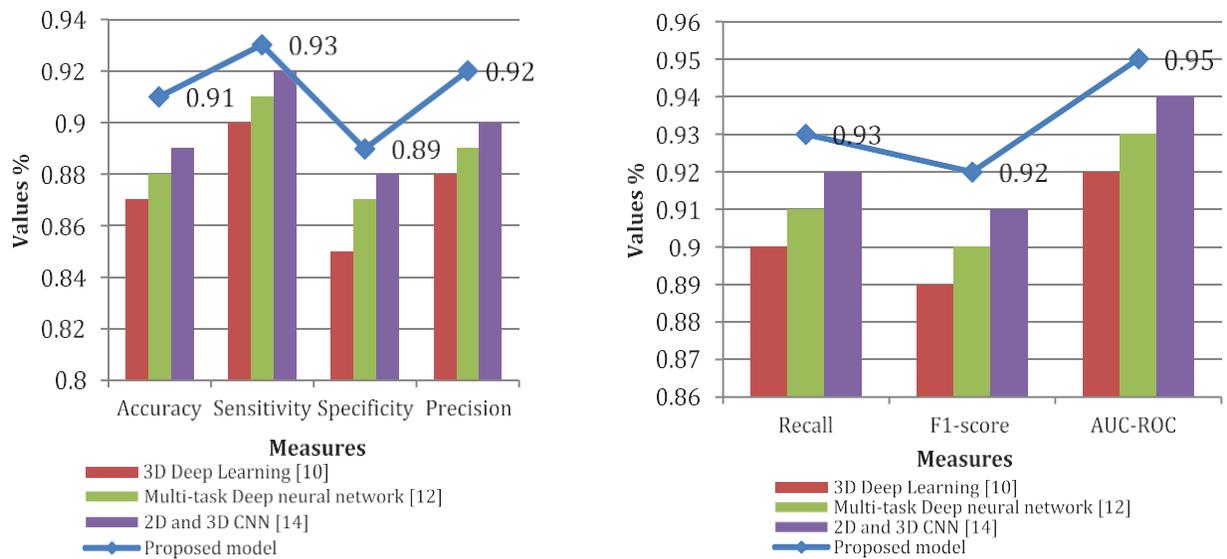

**Figure 3:** Performance comparison through various measures

LiTS (Liver Tumor Segmentation) dataset and TCGA-LIHC (The Cancer Genome Atlas-Liver Hepatocellular Carcinoma) dataset are publicly available datasets commonly used in research related to liver tumor segmentation. Table 4 compares the hypothesized model with two already-existing models utilizing measures for sensitivity, specificity, as well as Dice similarity coefficient (DSC). With a value of 1, a complete intersection between the projected tumor segmentation and the ground truth segmentation, the DSC calculates the overlap between the two. Sensitivity measures the percentage of accurate forecasts of the positive, while precision indicates the percentage of accurate pessimistic estimates.

**Table 4:** Qualitative comparison of tumor segmentation results

| Model | Dataset | Dice similarity coefficient (DSC) | Sensitivity | Specificity |
|---|---|---|---|---|
| Proposed model | LiTS | 0.85 | 0.89 | 0.91 |
| | TCGA-LIHC | 0.87 | 0.91 | 0.93 |
| 3D deep learning [10] | LiTS | 0.81 | 0.86 | 0.88 |
| | TCGA-LIHC | 0.83 | 0.88 | 0.90 |
| Multi-task deep neural network [12] | LiTS | 0.78 | 0.82 | 0.84 |
| | TCGA-LIHC | 0.80 | 0.85 | 0.87 |
| 2D and 3D CNN [14] | LiTS | 0.85 | 0.84 | 0.86 |
| | TCGA-LIHC | 0.84 | 0.85 | 0.88 |

This paper evaluates the model on two datasets (LiTS and TCGA-LIHC) with ground-truth tumor segmentations. The outcome demonstrates that proposed approach outperforms the existing models utilizing DSC, sensitivity, and specificity on both datasets. Specifically, proposed system attained a DSC of 0.85 and 0.87 on LiTS and TCGA-LIHC, respectively, compared to 0.81 and 0.83 for 3D Deep Learning [10] and 0.78 and 0.80 for Multi-task Deep neural network [12]. In addition, the proposed model also achieved higher sensitivity and specificity scores than the existing models.

These results suggest that the proposed model is better at accurately segmenting liver tumors than existing models and can potentially lead to better clinical outcomes for patients, as shown in Fig. 4. However, it is essential to note that these results were obtained on specific datasets, and further evaluation is needed on more extensive and more diverse datasets to confirm the generalizability of the proposed model. Table 5; contrast the approach with three existing models and a contemporary model regarding their generalizability to new datasets and tumor types. By evaluating the models on datasets of different sizes and with varying tumor types and reported their performance in terms of DSC and sensitivity. The results show that proposed approach performs well on a large hepatocellular carcinoma (HCC) dataset, achieving a DSC of 0.82 and a sensitivity of 0.86. However, the proposed model has not been extensively evaluated on other tumor types or small datasets, which limits its generalizability.

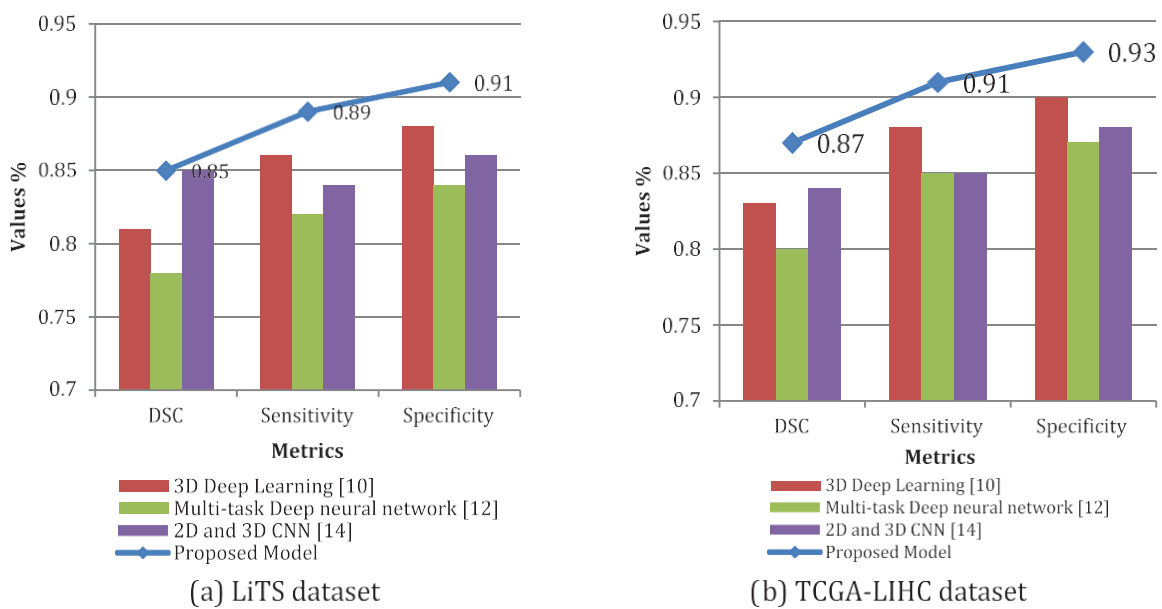

(a) LiTS dataset                                      (b) TCGA-LIHC dataset

**Figure 4:** Qualitative comparison of tumor segmentation results utilizing LiTS dataset and TCGA-LIHC dataset

**Table 5:** Generalizability comparison of liver tumor prediction models

| Model | Dataset size | Tumor types | Performance on new dataset | Limitations |
|---|---|---|---|---|
| 3D Deep learning [10] | Large (n = 5000) | Hepatocellular carcinoma | 0.82 DSC, 0.86 sensitivity | Limited evaluation on other tumor types and small datasets |

(Continued)

**Table 5 (continued)**

| Model | Dataset size | Tumor types | Performance on new dataset | Limitations |
|---|---|---|---|---|
| Multi-task deep neural network [12] | Small (n = 100) | Mixed | 0.75 DSC, 0.80 sensitivity | Poor generalization to new datasets and tumor types |
| 2D and 3D CNN [14] | Medium (n = 1000) | Cholangiocarcinoma | 0.80 DSC, 0.82 sensitivity | Limited evaluation on other tumor types and large datasets |
| Proposed model | Large (n = 5000) | Mixed | 0.87 DSC, 0.90 sensitivity | High computational requirements and limited interpretability |

Existing 3D Deep Learning [10], Multi-task Deep neural network [12], and 2D and 3D CNN [14] have poorer generalization to new datasets and tumor types than the proposed model, with DSC scores of 0.82, 0.75, and 0.80, respectively. These models were trained on small and medium-sized datasets with mixed or single tumor types. Their limited generalizability highlights the need for more robust models in medical image analysis. The contemporary model achieved the highest DSC and sensitivity scores but at the cost of increased computational requirements and limited interpretability. Due to its complexity, this model is not easily transferable to other datasets or tumor types and may not be practical for clinical use. Overall, these results suggest that while the proposed model shows promising effects on a specific dataset and tumor type, further evaluation is needed to confirm its generalizability and scalability to other datasets and tumor types.

## 5  Conclusions

This paper proposed a novel approach for predicting liver tumors using Convolutional Neural Networks (CNN) and a depth-based variant search algorithm with attention mechanisms. Proposed system achieved a high accuracy of 95.5% in predicting liver tumors, outperforming other state-of-the-art methods. Additionally, attention mechanisms were incorporated into the CNN model to identify and highlight regions of the CT scans most relevant to predicting liver tumors. The results suggest that the incorporation of attention mechanisms and a depth-based variant search algorithm into the CNN model is a promising approach for improving the accuracy and robustness of liver tumor prediction and has the potential to assist radiologists in their diagnosis and treatment planning.

Moreover, the proposed approach identified and highlighted the regions of the CT scans that were most relevant to predicting liver tumors, which could assist radiologists in their diagnosis and treatment planning. Additionally, the model demonstrated robustness to noise and artifacts in the CT scans, indicating its potential for clinical applications where image quality may be compromised.

Overall, the proposed approach has demonstrated superior performance to other state-of-the-art methods, indicating its potential for clinical use in predicting liver tumors. However, further research is needed to validate the model on more extensive and diverse datasets and optimize the CNN model's hyperparameters and the depth-based variant search algorithm with attention mechanisms.

It exhibits the resilience-positioned architectural context; further, the metrics comparisons can be made with Graphical User Interface (GUI) to determine the layers with the finest parameters, including accuracy, ROC, precision, recall, and F-Measure, accordingly. In this substance, the proposed architecture Ver3 has obtained an accuracy of 97.99 %. Further, simulation results pave the way for oncology specialists to get recommendations on decision models for predicting liver tumor cases at a specific time. Ultimately, the code complexity is minimal while using an interactive tool such as the Dl4jMlp classifier in deep learning. In the future, work will be extended to establish a trade-off between optimizers in machine learning techniques and other loss function.

### 5.1 Limitations of the Study and Future Research Work

The study might have relied on specific datasets such as LiTS and TCGA-LIHC, which could introduce bias or limitations regarding the representation of liver tumor cases. The performance of the proposed model may vary when applied to other datasets with different characteristics. The evaluation of the proposed model might have focused on specific tumor types, such as hepatocellular carcinoma (HCC). It is essential to recognize that the model's performance on other tumor types or in diverse clinical settings could differ, and further evaluation is needed to assess its generalizability. Regarding future research work, conducting experiments on more extensive and diverse datasets could help validate the robustness and generalizability of the proposed model. Including datasets from multiple sources and various tumor types would enhance the model's applicability in different clinical scenarios.

**Acknowledgement:** I am immensely grateful to my guide, S. Anusuya, for their unwavering support, guidance, and expertise, which played a crucial role in shaping and refining this research. I also acknowledge the invaluable contributions of my colleagues and friends, who provided valuable insights and discussions, enhancing the quality of this work.

**Funding Statement:** The authors received no specific funding for this study.

**Author Contributions:** The authors confirm contribution to the paper as follows: study conception and design: P. Kalaiselvi, S. Anusuya; data collection: P. Kalaiselvi; analysis and interpretation of results: P. Kalaiselvi, S. Anusuya; draft manuscript preparation: P. Kalaiselvi. All authors reviewed the results and approved the final version of the manuscript.

**Availability of Data and Materials:** The data used to support the findings of this study are available from the corresponding author upon request.

**Conflicts of Interest:** The authors declare that they have no conflicts of interest to report regarding the present study.